# High-throughput screening for modulators of cellular contractile force.

Chan Young Park[1], Enhua H. Zhou[1], Dhananjay Tambe[1], Bohao Chen[2], Tera Lavoie[2], Maria Dowell[2], Anton Simeonov[3], David J. Maloney[3], Aleksandar Marinkovic[1], Daniel J. Tschumperlin[1], Stephanie Burger[1], Matthew Frykenberg[1], James P. Butler[1], W. Daniel Stamer[4], Mark Johnson[5], Julian Solway[2], Jeffrey J. Fredberg[1], Ramaswamy Krishnan[6]

1. 665 Huntington Avenue, Molecular and Integrative Physiological Sciences, Harvard School of Public Health, Boston, MA 02115.
2. 5841 S. Maryland Avenue, Departments of Medicine, University of Chicago, Chicago, IL 60637
3. 6701 Democracy Boulevard, National Center for Advancing Translational Sciences, National Institutes of Health, Bethesda, MD 20892
4. 2351 Erwin Road, Duke Eye Center, Duke University, Durham, NC 27710
5. 2145 Sheridan Road, Department of Biomedical Engineering, Northwestern University, Evanston, IL 60208
6. 99 Brookline Avenue, Center for Vascular Biology Research, Beth Israel Deaconess Medical Center, Boston, MA 02215

Corresponding author:
   Ramaswamy Krishnan, Ph.D.
   99 Brookline Avenue, Center for Vascular Biology Research, Beth Israel Deaconess Medical Center, Boston, MA 02215
   617-667-2572
   Rkrishn2@bidmc.harvard.edu

**Abstract**

**When cellular contractile forces are central to pathophysiology, these forces comprise a logical target of therapy. Nevertheless, existing high-throughput screens are limited to upstream signaling intermediates with poorly defined relationship to such a physiological endpoint. Using cellular force as the target, here we screened libraries to identify novel drug candidates in the case of human airway smooth muscle cells in the context of asthma, and also in the case of Schlemm's canal endothelial cells in the context of glaucoma. This approach identified several drug candidates for both asthma and glaucoma. We attained rates of 1000 compounds per screening day, thus establishing a force-based cellular platform for high-throughput drug discovery.**

**Main Text**

In many organs and tissues, cellular contractile forces play a central role in physiology and pathophysiology.  As such, modulation of cellular contractile forces is often the main therapeutic strategy.  Commonplace examples are cardiac inotropes for cardiomyocytes (1), bronchodilators for airway smooth muscle cells (2), vasodilators for vascular smooth muscle cells (3), and relaxants for skeletal muscle cells (4).  Cellular contractile forces are also important in metastasis and cancer cell invasion (5, 6).  In each of these instances there clearly exist urgent unmet therapeutic needs (7-9).  Nevertheless, it has not previously been practical to use measurements of cellular contractile forces themselves as a primary read-out in high-throughput drug discovery.  Instead, currently available high-throughput screening technologies have been limited to measurements of surrogates for contraction itself, including upstream effectors such as intracellular messengers, binding affinity assays against specific cell surface receptors or other moieties, protein expression and protein relocation, or morphological changes (10-13).

The strength of these existing high-throughput approaches is that they are remarkably fast, but the weakness is that they stop short of incorporating and directly evaluating the main therapeutic target – cellular contractile force.  Depending upon the assay, therefore, not only are certain drug candidates potentially missed, but also many of the corresponding hits might be found subsequently to have little or no impact on contractile force.  Necessarily, efficacy and validation of these hits can only be established independently using old-fashioned low throughput methods such as contractility measured in the isolated muscle strip (14) or reactivity measured in the living organism (15).  Cellular deformability has been proposed as a high-throughput basis for cell screening in the context of cancer (16), malaria (17) and malignant pleural effusions (18) but these assays are limited to floating cells and are insensitive to depolymerization of actin, inhibition of myosin (19), or modulation of adhesion proteins.  In anchorage-dependent cells, and especially when contractile force is of concern, such assays are inapplicable.  In the context of physiological and organ systems modeling, *in-vitro* tissue constructs and organ-on-chip technologies are also promising (20-22) but, compared to the approach described below, are considerably more complex and thus less well suited for high levels of screening throughput.  Overall, many potential drugs have been found via current high-throughput assays, but the majority of new molecular entities approved by US Food and Drug Administration (FDA) continue to be discovered via traditional phenotypic assays (23).  Moreover, because 50% of drug candidates currently fail in phase II clinical trials (24), it has been suggested that decreased failure rates and reduced development costs might be attained if disease-relevant endpoints were brought into drug discovery at an earlier stage (25).  To fill this gap, we describe here a simple new assay based upon straightforward measurement of cellular contractile force itself, which serves as the targeted physiological endpoint.

Contractile force screening (CFS) is based upon Fourier-transform traction microscopy (26-29), which we adapted to 96-well plates (30).  In each well, polyacrylamide gel surfaces were labeled with fluorescent markers (31), functionalized with collagen (32), and covered with cells grown to near confluence (27-29, 33) (Fig. 1A-B, Supplementary Fig. 1).  Using an automated fluorescence microscope, we quantified in each well the average (i.e. root mean square) cellular contractile forces before and after adding drugs.  Drug effects were quantified as the 'force response ratio', namely, the contractile force before versus after drug addition (Fig. 1C).  For example, the average contractile force generated by cultured primary human airway smooth muscle (HASM) cells at baseline was 38 Pa (left column in Fig. 1C), and force did not change after adding vehicle alone (dimethyl sulfoxide (DMSO), 0.5%, final concentration); thus, the force response ratio for vehicle was close to 1.  The force response ratio for fetal bovine serum (FBS, 1%), which is known to increase contraction (34), was 1.77 (middle column in Fig. 1C), while that for the rho kinase inhibitor Y27632 (10 µM)(35), known to impair contraction, was 0.29 (right column in Fig. 1C).  Force response ratios revealed dose-dependent increases or decreases in cellular contractile force induced by FBS (Fig. 1D), Y27632, and the airway smooth muscle cell relaxant isoproterenol (Fig. 1E). Furthermore, using these force response ratios, CFS demonstrated that the robust Z'-factor was bigger than 0.6 (methods; Supplementary Fig. 3), thus confirming that the force response ratio provides a methodologically simple, physiologically relevant, and statistically valid index for identifying compounds that modulate cellular contractile forces.

To test the utility of CFS in the context of drug repurposing for use in asthma, we focused upon HASM cells. During an acute asthma attack, contractile forces generated by HASM cells act to constrict the airway and thus obstruct airflow. To dilate constricted airways, asthma patients use bronchodilators to reduce these contractile forces and thus allow the airway to open more fully, but currently available bronchodilator medications often fail to relax that muscle sufficiently, especially in severe asthma (2). To address this unmet need for more effective bronchodilator medications, we screened the Prestwick Chemical Library®, comprised of 1,120 drugs already approved by the FDA or European Medicines Agency (EMA), to identify which among these might be an unanticipated candidate to relax airway smooth muscle cells in asthma. Each datum plotted represents the average of quadruplicate measurements of a mixture of 4 drugs/well, with a concentration of 6.5µM for each in the initial screen (Fig. 2A, left panel); the middle panel is the histogram of all responses and the right panel is the rank-ordered response. Mixtures that modulated contractile force appreciably – termed positives – were later retested individually to identify the active drugs – termed hits. Most mixtures did not change cellular contractile forces appreciably. Several mixtures increased the force response ratio more than did FBS, and, more importantly, several mixtures decreased the response ratio as much as or more than control relaxant compounds. From 280 mixtures tested, we selected 15 mixtures as positives (shown in red in Fig. 2A) that were found to blunt contractile force appreciably. After retesting drugs individually, we found 15 hits; 9 were β-adrenergic receptor agonists and 3 (alprostadil (36, 37), ethaverine hydrochloride (38) and kaempferol (39, 40)) were already well-known as smooth muscle relaxants (Supplementary Table 1A). However, the HASM relaxant effects of three drugs were unexpected: Chicago sky blue, terconazole and levonordefrin (α-methylnorepinephrine). These findings confirm the ability of CFS to identify novel relaxants of airway smooth muscle.

To validate further the utility of CFS in a different disease context, we turned to human endothelial cells of Schlemm's canal (SC) in the context of drug repurposing for use in glaucoma, which remains a leading cause of blindness (41). All current drug treatments and surgical treatments for glaucoma target reduction of intraocular pressure, but many patients remain refractory to those treatments. Because excessive contraction of the SC cell has been implicated recently in the etiology of glaucoma (42, 43), we first tested human SC cells using the control drugs described above and found that they modulated contractile force of SC cells much as they did in HASM cells (Fig. 1F). Next, from the Prestwick Chemical Library®, 17 individual drugs were identified as hits that blunted SC cell contraction. One was a toxin (Sanguinarine), 9 were β-adrenergic receptor agonists already well-known as SC relaxants, and one was a vasodilator (Alprostadil, Supplementary Table 1B). Alprostadil was the most potent of these hits; further inspection revealed that alprostadil reduced contractile force of SC cells in a dose-dependent manner (supplementary Fig. 4A) and its efficacy was further validated in perfusion studies testing aqueous humor outflow function in enucleated mouse eyes (Supplementary Fig. 4B).

Having established the feasibility of using CFS in a small library like the Prestwick Chemical Library®, we turned next to larger libraries and questions of throughput. In a subset comprising 10,000 compounds selected randomly from the Chembridge DIVERset®, we set up each drug plate to contain 35 different compound mixtures and 5 controls, with each mixture consisting of 8 compounds. Since hits are few (Fig. 2), the probability of interactions between compounds or cancelling effects in any given well is small. For this screening, we used HASM cells to find novel candidate bronchodilators. With drug incubation time of 1 hr and using a single microscope, we screened all 10,000 compounds at the rate of 1,120 compounds per screening day (Supplementary Fig. 5). From this screen, we found 12 positives (Fig. 2B) and finally 2 hits that were closely related structurally. Further studies of these compounds revealed that contractile forces in HASM cells were reduced substantially, non-toxically, and in a dose-dependent manner (Supplementary Fig. 6).

In summary, Contractile Force Screening is a high throughput technology that directly addresses the physiologic target of interest, with an overall throughput on the order of at least a thousand compounds per microscope per day. Because this study did not employ automation or robotic handing, substantially higher levels of throughput should be readily attainable. CFS can thus fill an important methodological void in the middle ground between high-throughput but relatively non-physiological approaches on the one hand and physiological but low-throughput animal or tissue-based approaches on the other. As such, CFS has the potential to facilitate drug discovery and drug repurposing in any circumstance in which modulation of contractile force is the logical therapeutic target, including vascular and cardiac disease, pulmonary arterial hypertension, asthma, glaucoma, and metastatic and invasive disease.

**Methods:**

**Cell culture:** Primary human airway smooth muscle (HASM) cells were obtained from lungs unsuitable for transplantation, as previously described (32, 44). Passage 5-7 cells from 5 donors were used for assay development. Screening and secondary validations of both the Prestwick® and Chembridge DIVERset® libraries were restricted to passage 7 cells from one donor. 10,000 cells per well were seeded onto assay plates in a medium containing 10% fetal bovine serum (FBS). After 2 hours of incubation, the serum containing medium was replaced with serum free medium containing insulin (5.7µg/mL) and transferrin (5µg/mL) instead of FBS for an additional 48 hour prior to experimentation.

Human endothelial cells of the inner wall of Schlemm's canal (SC) were obtained from post-mortem human eyes provided by Midwest Eye Bank, NDRI or Life Legacy as previously described (43, 45); passage 6-7 from 1 donor were used. Approximately 3,200 to 6,400 cells per well were seeded onto assay plates in a medium containing 1% FBS. Cells were grown in this medium for 2 days and in serum free medium supplemented with insulin-transferrin-selenium for an additional 12 hours prior to experimentation.

**Preparation of drug mixtures**: From the Prestwick Chemical Library®, we screened 1,120 drugs. 4 drugs within the same column of the source plate were mixed together and distributed within drug plates as shown in Supplementary Figure 2. From the Chembridge DIVERSet Library®, we screened 10,000 compounds. 8 drugs within same column in the source plates were mixed together and distributed within drug plates as shown in Supplementary Figure 2.

**Perfusion of mouse eyes:** Perfusions were performed in enucleated eyes obtained from 11 C57BL/6 mice of either gender, aged 10 weeks to 7 months old at time of death. Enucleated eyes were stored in phosphate buffered saline at 4°C until perfusion, typically 1-3 hours. The perfusion method follows previously described techniques that we developed (46-50) with a few modifications. Briefly, each enucleated eye was affixed onto a post using cyanoacrylate glue; stabilizing the eye for cannulation of the anterior chamber. Eyes were cannulated with a 33-gauge beveled-tip needle (Nanofil; World Precision Instruments, Europe; Hitchin, UK) backfilled with 1 µM alprostadil or vehicle (ethanol, 1: 10,000). Housed in a humidified chamber, eyes were perfused in pairs and randomized for each perfusion as to whether drug or vehicle was perfused first. The needle was connected via pressure tubing to a glass syringe (25 µL; Hamilton GasTight, Reno, NV) that was mounted and controlled by motorized syringe pump (PHD Ultra; Harvard Apparatus, MA). Custom written

LabVIEW software (National Instruments Corp., Austin, TX) served to monitor intraocular pressure (IOP) (via in line 142PC01G pressure transducer; Honeywell, Columbus, OH) and control the flow rates delivered by the syringe pump into the eye to maintain a user-defined IOP (13). Both experimental and control eyes were initially held at 8mmHg for 45 minutes using a fluid reservoir to facilitate exposure of cells in the outflow pathway to drug (or vehicle). Subsequently, eyes were perfused at sequential pressure steps of 4, 8, 15 and 20 mmHg. Each pressure step was maintained for 20-30 minutes to obtain a minimum of 10 minutes of stable flow data, from which an average stable flow rate was calculated for each pressure step. Data from an individual eye was considered acceptable if a stable flow rate was achieved in at least 3 of the 4 pressure steps. Outflow facility was found following the general principle of the two-level constant pressure perfusion procedure introduced by Barany (51). Here we measured flow rate (Q) at four pressures (P), and then found the outflow facility using a regression analysis by fitting the data to the following relationship using SPSS:

$$Q = a_0 + a_1 * P + a_2 * P * DRUG$$

where DRUG=0 is the control case and DRUG=1 when alprostadil was applied. $a_1$ is the outflow facility for control eyes and $a_1 + a_2$ that for eyes after alprostadil application.

**Preparation of deformable substrates in 96-well plates:** Polyacrylamide based gel substrates were miniaturized in glass bottom 96-well plates using one of two methodologies. In the first method, each 96-well plate was treated with NaOH (6N in water) for 1hr followed by silane solution ((3-aminopropyl)trimethoxysilane, 10% in water) for an additional 1hr. Next, red fluorescent bead solution (1µm carboxylate-modified microspheres, Invitrogen, $2 \times 10^{-4}$% in water) was added to the wells and air-dried overnight. Dried glass surfaces were then treated with glutaraldehyde (0.25% in PBS) for 30min and further washed and dried. Acrylamide gels (5.5% acrylamide, 0.076% bisacrylamide, Young's modulus = 1.8kPa, thickness = 200 µm) were cast in each well using a custom-made gel caster (30) (Matrigen Life Technologies, CA). The gel surfaces were functionalized using sulfo-SANPAH (sulfosuccinimidyl-6-[4´-azido-2´-nitrophenylamino]hexanoate, 0.2mg/mL), coated with green beads (0.2µm sulfate microspheres, Invitrogen, $2 \times 10^{-3}$% in water) (31), coated with bovine collagen I (40µg/mL in PBS) (32) and were stored at 4°C (Fig. 1A).

In the second method, each 96-well plate was treated with silane (γ-methacryloxypropyltrimethoxysilane, 0.4% in water), and a first layer of acrylamide gel substrate (8% acrylamide, 0.1% bisacrylamide, and 0.4% acrylic acid N-hydroxysuccinimide ester (NHS)(52), Young's modulus ≈ 2.5 kPa, thickness ≈ 200 µm) was prepared using the custom-made gel caster described above. The casting procedure was repeated for a second layer of gel (53) with the same composition plus 2% vol/vol red beads (0.5µm carboxylate-modified microspheres, Invitrogen); this top layer was prepared exceedingly thin to promote fluorescent bead dispersion within a single horizontal plane. The gels were coated with bovine collagen I (10 µg/mL in PBS) (Supplementary Fig. 1).

**Measurements of contractile forces using Fourier-transform traction microscopy:** The 96-well plate was mounted within a heated chamber (37°C) upon a motorized stage and imaged using an inverted microscope (DMI 6000B, Leica Inc.). In each well, three images were obtained in quick succession: one phase contrast image of cells and a pair of fluorescent images of beads (Fig 1A). The image set was obtained before plating cells (*reference*), immediately prior to adding drugs (*baseline*), and 1 hr after drug addition (*treatment*). By comparing fluorescent images obtained during *baseline* or *treatment* with the corresponding image from *reference,* we computed the cell-exerted displacement field (26, 27). From the displacement field, we computed the contractile force (per unit area) using Fourier-transform traction microscopy (26, 27) modified to the case of cell monolayers (27-29, 52). This modified approach takes into consideration effects of finite gel thickness as well as force imbalances associated with the microscope field of view as we described previously (27, 28). From each force map (Fig 1B), we computed the root mean squared value to represent the averaged contractile force. Throughout the paper, we use the generic word "force" to mean the traction which is the contractile force (per unit area) that cells exert on their substrate.

**Evaluation of CFS:** Traditional Z'-factor (54) has been widely used as a quality metric for small-molecule screens but it has limitations such as its sensitivity on data distribution or outliers (55). Therefore alternative metrics have been suggested including the robust Z'-factor (56). The variability of cellular stiffness and contractile force in HASM cells are known to be very high (57, 58) and the distribution of the force response ratios are closer to a log-normal distribution (Fig. 2). Hence, we used here the robust Z'-factor as a quality metric of CFS. The robust Z'-factor is defined as below:

$$robust\ Z' = 1 - \frac{3(S1 + S2)}{|X1 - X2|}$$

where, *S1* and *S2* are the median absolute deviations of negative and positive controls and *X1* and *X2* are the medians of negative and positive controls. We used vehicle control (water, 0.01%) as the negative control and rho-kinase inhibitor, Y27632 (12μM) as the positive control. The control compounds were distributed equally in the 96 well-plate and examined for their effects on cellular contractile forces. From these measurements, we computed the robust Z'-factor and it was 0.605 (Supplementary Fig. 3).

**Acknowledgments:** The authors are grateful to Dr. Reynold Panettieri at University of Pennsylvania for providing primary HASM cells, to the Institute for Genomics and Systems Biology at University of Chicago for preparing the drug plates, and to the Institute of Chemistry and Cell Biology at Harvard Medical School for technical assistance. We gratefully acknowledge the critical comments of Emil Millet and Dr. Ajit Jadhav and the support provided by the National Institutes of Health grants R01EY019696, R01HL102373, R01HL107561, P01HL120839, and UH2HL123816.

**AUTHOR CONTRIBUTIONS**
C.Y.P, J.S, J.J.F, and R.K. conceived the contractile force screening. C.Y.P., E.H.Z., A.M., D.J.T., S.B., M.F., and J.J.F developed assay plates and screening protocols. C.Y.P., E.H.Z., S.B., and M.F. performed screenings and secondary validations. W.D.S isolated primary SC cells and performed mouse eye perfusion experiments. B.C., T.L., M.D. and J.S. isolated primary HASM cells. D.T. and J.P.B. guided software development. C.Y.P, E.H.Z, A.S., D.J.M., W.D.S., M.J., J.S., J.J.F., and R.K. guided data interpretation and analysis. All authors except A.M, S.B., M.F. contributed to manuscript writing. C.Y.P, J.J.F., and R.K. oversaw the project. All authors approved the manuscript.

**COMPETING FINANCIAL INTERESTS:** The authors declare no competing financial interests.

**References:**

1. Overgaard CB & Dzavik V (2008) Inotropes and vasopressors: review of physiology and clinical use in cardiovascular disease. *Circulation* 118(10):1047-1056.
2. Barnes PJ (2004) New drugs for asthma. *Nat Rev Drug Discov* 3(10):831-844.
3. Humbert M, Sitbon O, & Simonneau G (2004) Treatment of pulmonary arterial hypertension. *N Engl J Med* 351(14):1425-1436.
4. See S & Ginzburg R (2008) Choosing a skeletal muscle relaxant. *Am Fam Physician* 78(3):365-370.
5. Jonietz E (2012) Mechanics: The forces of cancer. *Nature* 491(7425):S56-57.
6. Wirtz D, Konstantopoulos K, & Searson PC (2011) The physics of cancer: the role of physical interactions and mechanical forces in metastasis. *Nat Rev Cancer* 11(7):512-522.
7. Hasenfuss G & Teerlink JR (2011) Cardiac inotropes: current agents and future directions. *Eur Heart J* 32(15):1838-1845.
8. Wechsler ME (2014) Getting Control of Uncontrolled Asthma. *Am J Med*.
9. Seferian A & Simonneau G (2013) Therapies for pulmonary arterial hypertension: where are we today, where do we go tomorrow? *Eur Respir Rev* 22(129):217-226.
10. Sams-Dodd F (2005) Target-based drug discovery: is something wrong? *Drug Discov Today* 10(2):139-147.
11. Brown D (2007) Unfinished business: target-based drug discovery. *Drug Discov Today* 12(23-24):1007-1012.
12. Zheng W, Thorne N, & McKew JC (2013) Phenotypic screens as a renewed approach for drug discovery. *Drug Discov Today* 18(21-22):1067-1073.
13. Zanella F, Lorens JB, & Link W (2010) High content screening: seeing is believing. *Trends Biotechnol* 28(5):237-245.
14. Fredberg JJ, Inouye DS, Mijailovich SM, & Butler JP (1999) Perturbed equilibrium of myosin binding in airway smooth muscle and its implications in bronchospasm. *Am J Respir Crit Care Med* 159(3):959-967.
15. Holmes AM, Solari R, & Holgate ST (2011) Animal models of asthma: value, limitations and opportunities for alternative approaches. *Drug Discov Today* 16(15-16):659-670.
16. Guck J, et al. (2005) Optical deformability as an inherent cell marker for testing malignant transformation and metastatic competence. *Biophys J* 88(5):3689-3698.
17. Mauritz JM, et al. (2010) Detection of Plasmodium falciparum-infected red blood cells by optical stretching. *J Biomed Opt* 15(3):030517.
18. Tse HT, et al. (2013) Quantitative diagnosis of malignant pleural effusions by single-cell mechanophenotyping. *Sci Transl Med* 5(212):212ra163.
19. Gossett DR, et al. (2012) Hydrodynamic stretching of single cells for large population mechanical phenotyping. *Proc Natl Acad Sci U S A* 109(20):7630-7635.
20. Legant WR, et al. (2009) Microfabricated tissue gauges to measure and manipulate forces from 3D microtissues. *Proc Natl Acad Sci U S A* 106(25):10097-10102.
21. Nesmith AP, Agarwal A, McCain ML, & Parker KK (2014) Human airway musculature on a chip: an in vitro model of allergic asthmatic bronchoconstriction and bronchodilation. *Lab Chip*.
22. Huh D, et al. (2010) Reconstituting organ-level lung functions on a chip. *Science* 328(5986):1662-1668.
23. Swinney DC & Anthony J (2011) How were new medicines discovered? *Nat Rev Drug Discov* 10(7):507-519.
24. Paul SM, et al. (2010) How to improve R&D productivity: the pharmaceutical industry's grand challenge. *Nat Rev Drug Discov* 9(3):203-214.
25. Lang P, Yeow K, Nichols A, & Scheer A (2006) Cellular imaging in drug discovery. *Nat Rev Drug Discov* 5(4):343-356.
26. Butler JP, Tolic-Norrelykke IM, Fabry B, & Fredberg JJ (2002) Traction fields, moments, and strain energy that cells exert on their surroundings. *Am J Physiol Cell Physiol* 282(3):C595-605.
27. Trepat X, et al. (2009) Physical forces during collective cell migration. *Nat Phys* 5(6):426-430.
28. Tambe DT, et al. (2011) Collective cell guidance by cooperative intercellular forces. *Nature Materials* 10(6):469-475.
29. Kim JH, et al. (2013) Propulsion and navigation within the advancing monolayer sheet. *Nature Materials* 12(9):856-863.

30. Mih JD, et al. (2011) A multiwell platform for studying stiffness-dependent cell biology. *PLoS One* 6(5):e19929.
31. Marinkovic A, Mih JD, Park JA, Liu F, & Tschumperlin DJ (2012) Improved throughput traction microscopy reveals pivotal role for matrix stiffness in fibroblast contractility and TGF-beta responsiveness. *Am J Physiol Lung Cell Mol Physiol* 303(3):L169-180.
32. Park CY, et al. (2010) Mapping the cytoskeletal prestress. *Am J Physiol Cell Physiol* 298(5):C1245-1252.
33. Krishnan R, et al. (2011) Substrate stiffening promotes endothelial monolayer disruption through enhanced physical forces. *Am J Physiol Cell Physiol* 300(1):C146-154.
34. Abdullah NA, et al. (1994) Contraction and depolarization induced by fetal bovine serum in airway smooth muscle. *Am J Physiol* 266(5 Pt 1):L528-535.
35. Uehata M, et al. (1997) Calcium sensitization of smooth muscle mediated by a Rho-associated protein kinase in hypertension. *Nature* 389(6654):990-994.
36. Sweatman WJ & Collier HO (1968) Effects of prostaglandins on human bronchial muscle. *Nature* 217(5123):69.
37. Wajima Z, et al. (2003) Intravenous alprostadil, an analog of prostaglandin E1, prevents thiamylal-fentanyl-induced bronchoconstriction in humans. *Anesth Analg* 97(2):456-460, table of contents.
38. Oswald WJ & Baeder DH (1975) Pharmacology of ethaverine HC1: human and animal studies. *South Med J* 68(12):1481-1484.
39. Leal LK, et al. (2006) Mechanisms underlying the relaxation induced by isokaempferide from Amburana cearensis in the guinea-pig isolated trachea. *Life Sci* 79(1):98-104.
40. Townsend EA & Emala CW, Sr. (2013) Quercetin acutely relaxes airway smooth muscle and potentiates beta-agonist-induced relaxation via dual phosphodiesterase inhibition of PLCbeta and PDE4. *Am J Physiol Lung Cell Mol Physiol* 305(5):L396-403.
41. Kahook M, Schuman JS, & Epstein DL (2013) Chandler and Grant's Glaucoma. *Slack Incorporated; 5th edition*
42. Overby DR, et al. (in press) Altered mechanobiology of Schlemm's canal endothelial cells in glaucoma. *Proc Natl Acad Sci U S A*.
43. Zhou EH, et al. (2012) Mechanical responsiveness of the endothelial cell of Schlemm's canal: scope, variability and its potential role in controlling aqueous humour outflow. *J R Soc Interface* 9(71):1144-1155.
44. Panettieri RA, Jr. (2001) Isolation and culture of human airway smooth muscle cells. *Methods Mol Med* 56:155-160.
45. Stamer WD, Roberts BC, Howell DN, & Epstein DL (1998) Isolation, culture, and characterization of endothelial cells from Schlemm's canal. *Invest Ophthalmol Vis Sci* 39(10):1804-1812.
46. Lei Y, Overby DR, Boussommier-Calleja A, Stamer WD, & Ethier CR (2011) Outflow physiology of the mouse eye: pressure dependence and washout. *Investigative ophthalmology & visual science* 52(3):1865-1871.
47. Boussommier-Calleja A, et al. (2012) Pharmacologic Manipulation of Conventional Outflow Facility in Ex Vivo Mouse Eyes. *Investigative ophthalmology & visual science* 53(9):5838-5845.
48. Boussommier-Calleja A & Overby DR (2013) The influence of genetic background on conventional outflow facility in mice. *Investigative ophthalmology & visual science* 54(13):8251-8258.
49. Stamer WD, Lei Y, Boussommier-Calleja A, Overby DR, & Ethier CR (2011) eNOS, a pressure-dependent regulator of intraocular pressure. *Investigative ophthalmology & visual science* 52(13):9438-9444.
50. Rogers ME, et al. (2013) Pigment epithelium-derived factor decreases outflow facility. *Investigative ophthalmology & visual science* 54(10):6655-6661.
51. Bárány EH (1964) Simultaneous measurement of changing intraocular pressure and outflow facility in the vervet monkey by constant pressure infusion. *Investigative Ophthalmology and Visual Science* 3:135-143.
52. Serra-Picamal X, et al. (2012) Mechanical waves during tissue expansion. *Nature Physics* 8(8):628-634.
53. Bridgman PC, Dave S, Asnes CF, Tullio AN, & Adelstein RS (2001) Myosin IIB is required for growth cone motility. *J Neurosci* 21(16):6159-6169.
54. Zhang JH, Chung TD, & Oldenburg KR (1999) A Simple Statistical Parameter for Use in Evaluation and Validation of High Throughput Screening Assays. *J Biomol Screen* 4(2):67-73.
55. Birmingham A, et al. (2009) Statistical methods for analysis of high-throughput RNA interference screens. *Nat Methods* 6(8):569-575.
56. Zhang XD (2007) A pair of new statistical parameters for quality control in RNA interference high-throughput screening assays. *Genomics* 89(4):552-561.

57. Fabry B*, et al.* (2001) Selected contribution: time course and heterogeneity of contractile responses in cultured human airway smooth muscle cells. *Journal of applied physiology* 91(2):986-994.
58. Krishnan R*, et al.* (2009) Reinforcement versus fluidization in cytoskeletal mechanoresponsiveness. *PLoS ONE* 4(5):e5486.
59. Iversen PW*, et al.* (2004) HTS Assay Validation. *Assay Guidance Manual*, eds Sittampalam GS, Gal-Edd N, Arkin M, Auld D, Austin C, Bejcek B, Glicksman M, Inglese J, Lemmon V, Li Z*, et al.*Bethesda (MD)).

**Figure Legends**

**Figure 1: Contractile force screening (CFS).** (A) Acrylamide-based hydrogels were miniaturized in glass bottom 96-well plates. (B) In each well, three images were obtained in quick succession: one phase contrast image of cells and a pair of fluorescent images of beads. (C) Maps of cellular contractile force (per unit area) before (top) and 1 hr after addition of drugs (bottom) together with the average magnitude indicated in the lower-left corner. Drug effects were quantified as the force response ratio, namely, the average contractile force before versus after drug addition. The dose-dependent force response ratio of (D-E) human airway smooth muscle (HASM) cells, and, (F) Schlemm's canal (SC) endothelial cells. Plotted in D and E are the average values ± SD and in F the average values ± SEM.

**Figure 2: CFS for novel bronchodilatory drugs.** (A) Each datum corresponds to the force response ratio for each mixture from (A) the Prestwick® chemical library, or, (B) the Chembridge DIVERset® library. Using the green dotted line as a cut-off, we selected for further evaluation the mixtures with the greatest relaxant effect, shown as red vertical lines. Shown on the far right in the left panel are the response ratio for 4 controls; DMSO, FBS, isoproterenol, and Y27632. The middle panel is the histogram with corresponding cut-off line and the right panel is the rank-ordered response. A few mixtures in top ranks were disregarded based on variability within quadruplicate measurements.

**Figure 1**

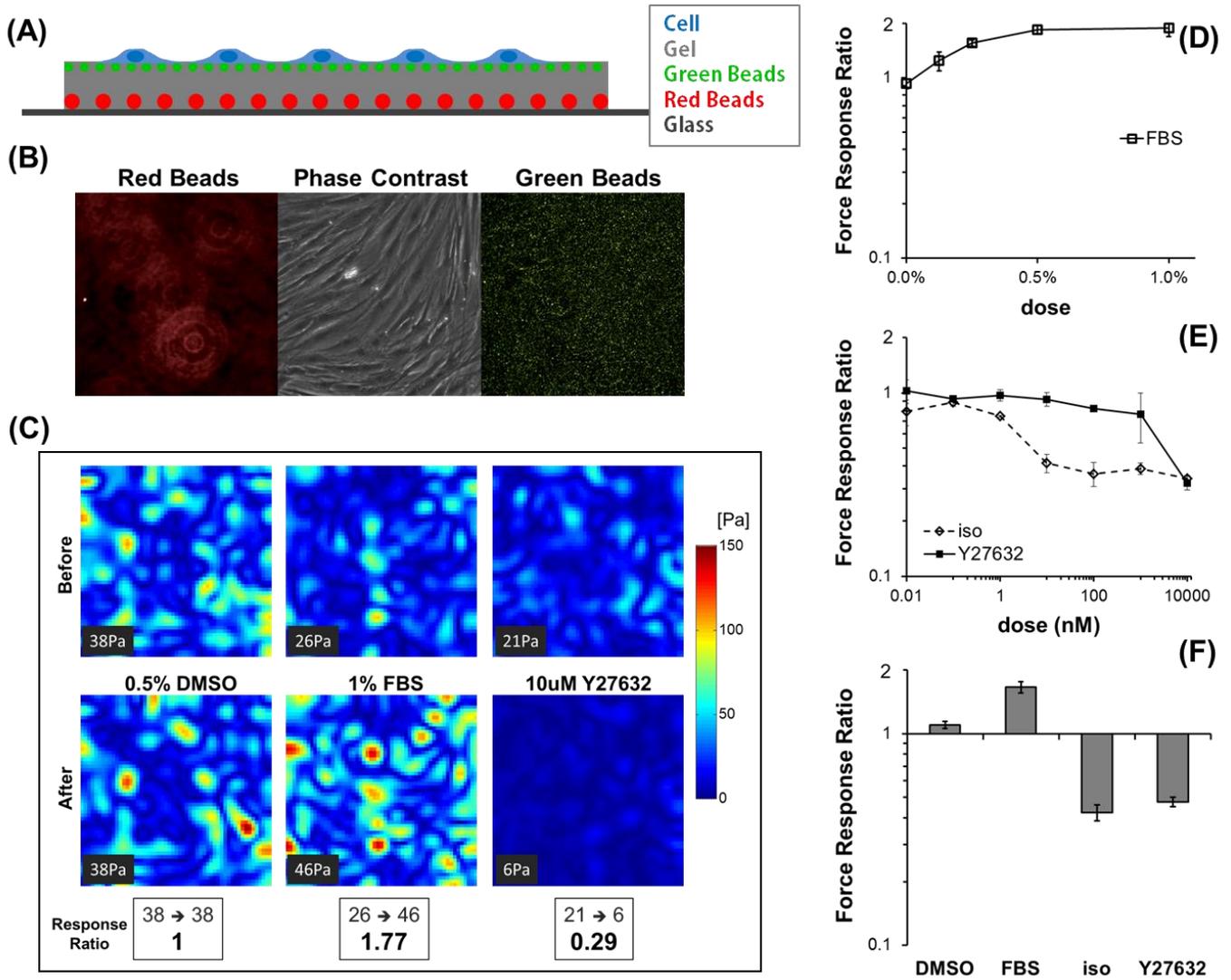

Figure 2

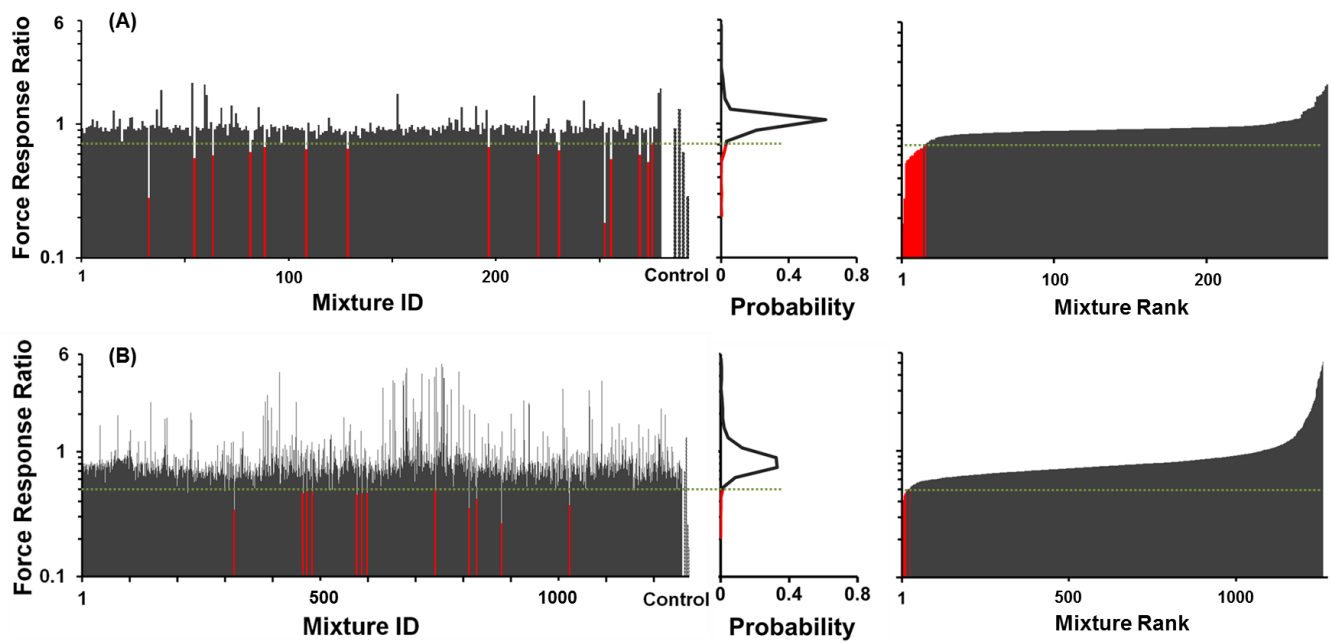

**Supplementary Figure 1: Preparation of deformable substrates in 96-well plates; an alternative method.**
Acrylamide-based hydrogels (Young's modulus ≈ 2.5kPa, thickness ≈ 200μm) were miniaturized in glass bottom 96-well plates using a multi-layered approach. Both layers had the same chemical composition with the exception of 2% wt/vol of red beads (0.5μm carboxylate-modified microspheres, Invitrogen) added to the top layer. The top layer was prepared exceedingly thin to promote bead dispersion within a single horizontal plane.

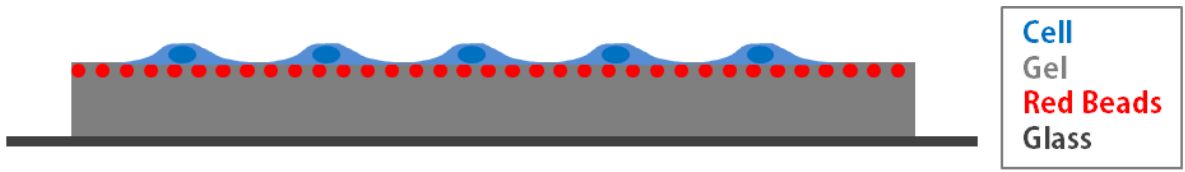

**Supplementary Figure 2: Layout of drug plate.** Each plate contained 35 different mixtures (1 to 35), vehicle control (0.5% DMSO), and three control drugs (1%FBS, 10μM or 25μM isoproterenol, and 10μM or 25μM Y27632).

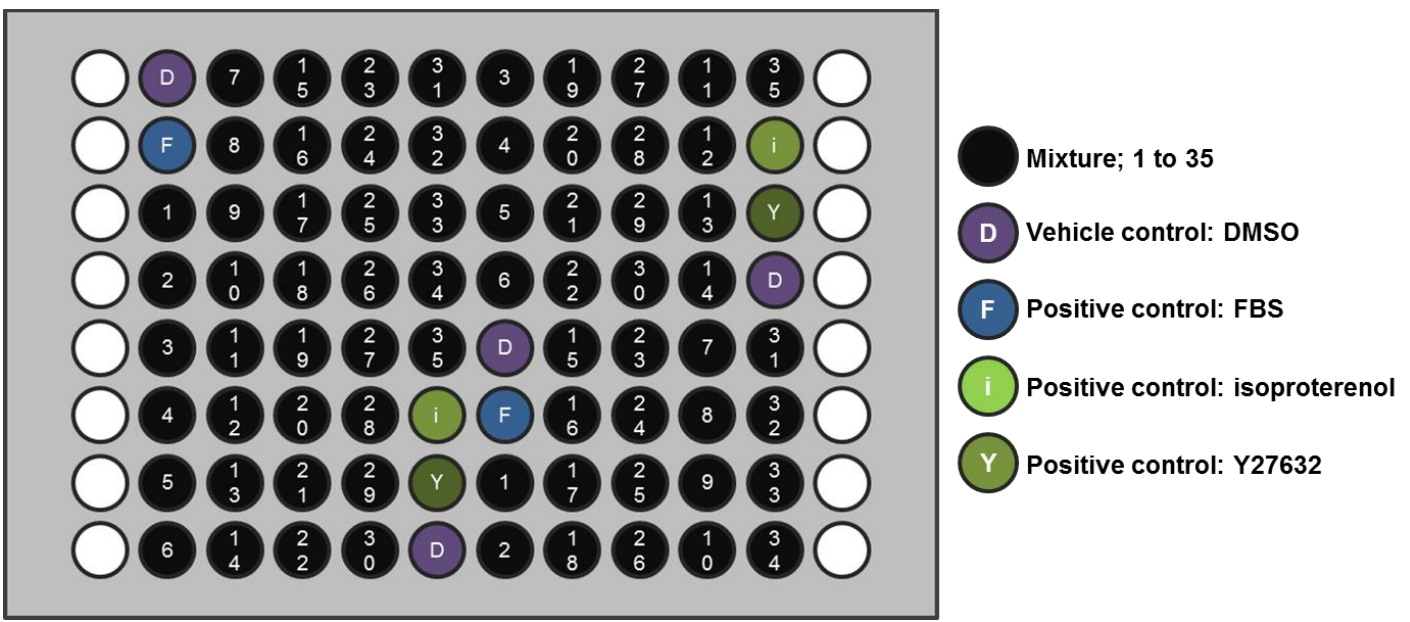

**Supplementary Figure 3: Force response ratio is a high-quality screening endpoint.** We assessed the high-throughput screening quality of CFS using the robust Z'-factor value (55, 56). The value was found to be 0.605 which exceeds the recommended value of 0.4 for a good quality HTS screen (59).

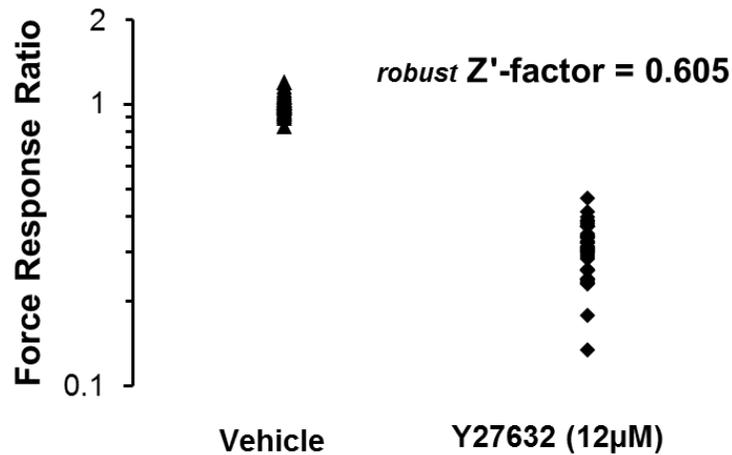

**Supplementary Figure 4: Alprostadil (prostaglandin E1) relaxes SC cells and promotes outflow function in enucleated mouse eyes.** (A) Alprostadil reduced SC contractile forces in a dose dependent fashion with greater potency than isoproterenol at a concentration less than 10µM. (B) 1µM alprostadil increased outflow facility (inverse of flow resistance) during perfusion of enucleated mouse eyes. ($p < 2x10^{-5}$, n=11 pairs). Plotted are the average ± SEM.

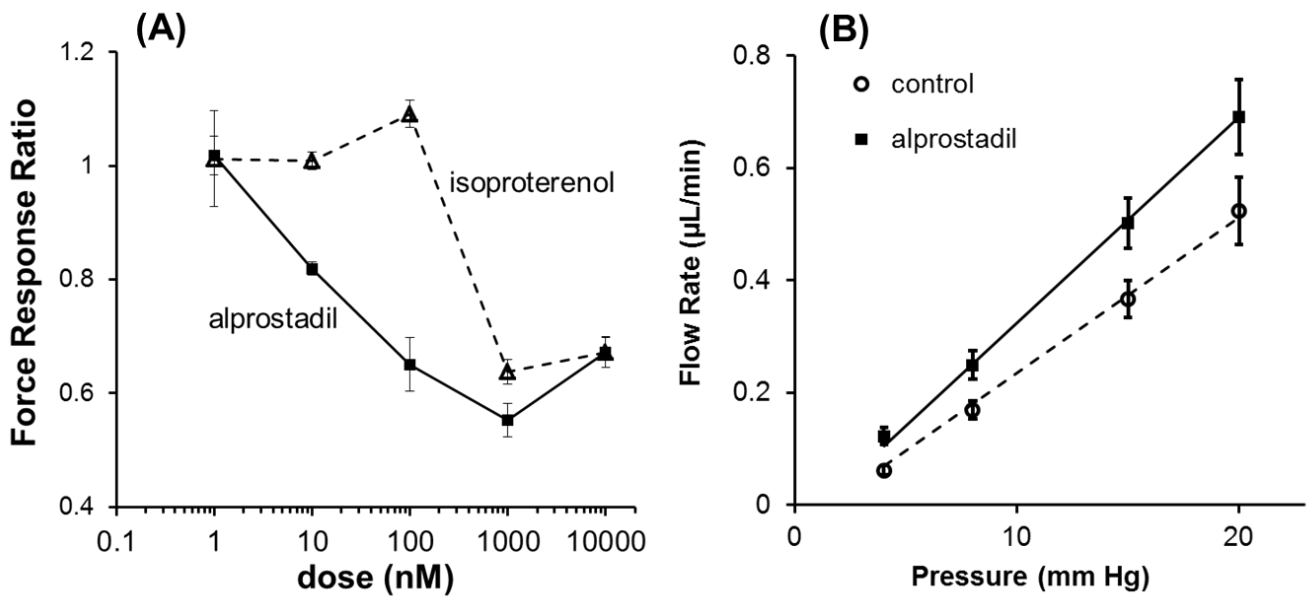

**Supplementary Figure 5: High-throughput implementation of CFS.** Using the Chembridge Diverset® library, we confirmed CFS throughput by screening 10,000 compounds within 4 weeks. The estimation is based on 96-well based assay plates, quadruplicate measurements, and 1hr drug incubation.

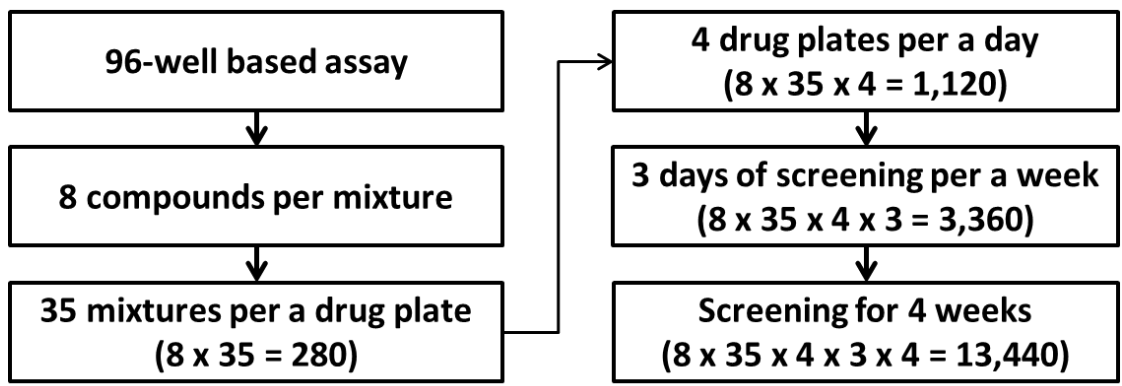

**Supplementary Figure 6: Secondary validation for novel bronchodilator hits identified by CFS.** (A) The 2 hits found from the Chembridge Diverset® did not affect cell viability depicted as the fluorescence intensity measured using a standard alamar Blue® assay (n=8 wells per compound per dose; 1% triton-X and 0.5% DMSO were used as negative and vehicle controls, respectively). (B) Both hit compounds reduced cellular contractile force in a dose dependent manner (n=8 wells per compound per dose). Plotted are the average ± SD.

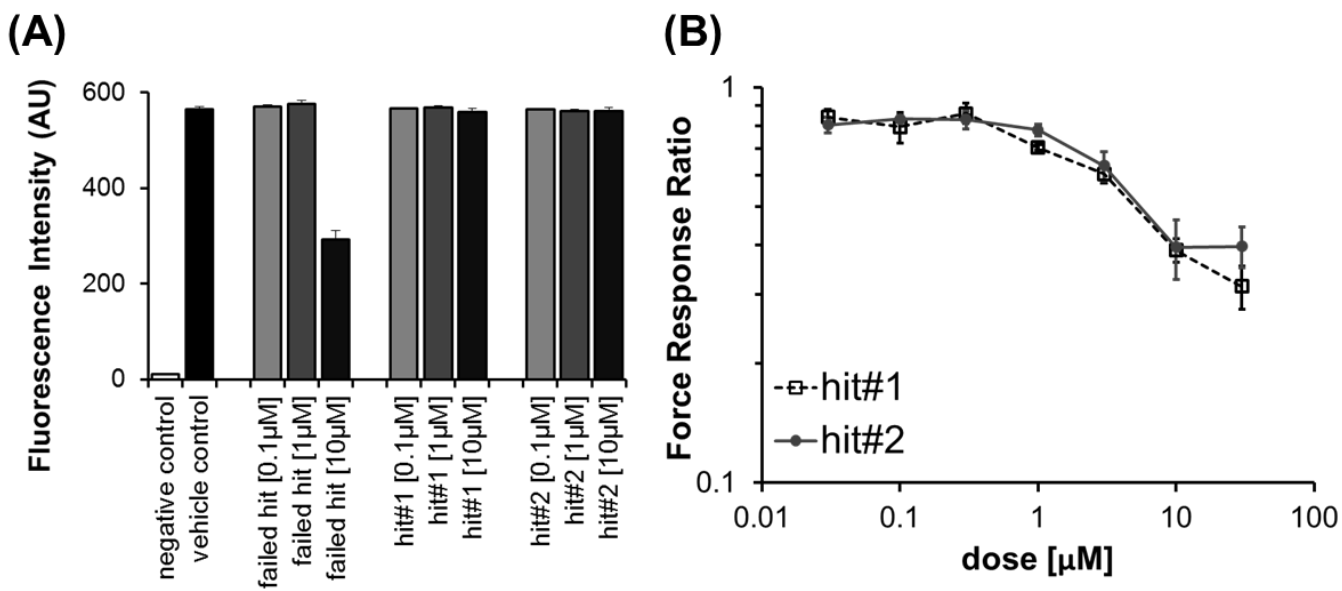

**Supplementary Table 1:** (A) From the screening of the Prestwick library, we identified 15 individual drugs that relaxed HASM cells. (B) From the screening of the Prestwick library, we identified 17 individual drugs that relaxed SC cells.

**Supplementary Table 1A:**

| Chemical name | Effects |
|---|---|
| Salbutamol | β2-adrenergic receptor agonist |
| Fenoterol hydrobromide | |
| Terbutaline hemisulfate | |
| Clenbuterol hydrochloride | |
| Ritodrine hydrochloride | |
| Metaproterenol sulfate | |
| Isoetharine mesylate salt | |
| (+)-Isoproterenol (+)-bitartrate salt | |
| (-)-Isoproterenol hydrochloride | |
| Alprostadil | Known smooth muscle relaxant |
| Ethaverine hydrochloride | |
| Kaempferol | |
| Chicago sky blue 6B | Inhibitor of L-glutamate uptake |
| Terconazole | Anti-fungal medication |
| Levonordefrin | Vasoconstrictor |

**Supplementary Table 1B:**

| Chemical name | Effects |
|---|---|
| Salbutamol | β2-adrenergic receptor agonist |
| Fenoterol hydrobromide | |
| Terbutaline hemisulfate | |
| Clenbuterol hydrochloride | |
| Ritodrine hydrochloride | |
| Metaproterenol sulfate | |
| Isoetharine mesylate salt | |
| (+)-Isoproterenol (+)-bitartrate salt | |
| (-)-Isoproterenol hydrochloride | |
| Pepstatin A | Inhibitor of aspartyl proteases, inhibits phosphorylation of ERK |
| Cyclosporin A | Immunosuppressant drug in organ transplantation |
| Clioquinol | Antifungal drug and antiprotozoal drug |
| Chicago sky blue 6B | Inhibitor of L-glutamate uptake |
| Terconazole | Anti-fungal medication |
| Levonordefrin | Topical nasal decongestant and vasoconstrictor in dentistry |
| Sanguinarine | Toxin |
| Alprostadil | Prostaglandin E1, known bronchodilator and vasodilator |